\documentstyle[ltwol,epsfig]{article}

\begin{document}

\title{SUMMARY AND OUTLOOK }

\author{R. D. Peccei}

\address{Department of Physics and Astronomy, University of California at Los Angeles\\Los Angeles, California 90095-1547}

\twocolumn[\maketitle\abstracts{ This summary is organized into four parts. In the first section (News) I discuss the SuperKamiokande results on atmospheric neutrino oscillations, as well as recent results from cosmology. The second section (Refinements) focuses on electroweak tests, recent results in the flavor sector and in probing QCD, as well as searches for new particles. The third section (Mysteries) discusses issues associated with neutrino masses and mixings in more depth. Finally, in the last section (Hopes) I reflect on both the short and long term future of the field.}]

\section{Introduction}

Giving a summary talk at a major conference is always a tricky business.
The approach most often taken is to focus on a few main points, discussing
each at some depth.  Rather foolishly, perhaps, I decided to take a
different tack for the Vancouver Conference and attempted to cover,
albeit at a somewhat superficial level, all the variegated and multifaceted
advances in the field presented at ICHEP 98.  The result was pleasing, at
least to me, for it demonstrated (once again) the enormous intellectual ferment
that particle physics can generate.  Although the Standard Model reigns
supreme, it is clear that our field is alive and well with plenty of hints
(and some evidence) of yet more exciting and far reaching discoveries nearby.
More importantly, the technical virtuosity, drive and imagination
displayed by the younger members of our profession at the meeting provided
the clearest assurance that particle physics will continue to have a
bright future.  Irrespective of what the ``new physics" may be, we will
find it!

I organized my talk into four parts:  News, Refinements, Mysteries and
Hopes.  In the first section, I concentrated on two fast breaking topics:
the evidence for neutrino oscillations coming from SuperKamiokande; and new
cosmological evidence pointing towards a Universe where the matter energy
density is less than the critical closing density.  The Refinements section
covered the bulk of the material presented at the Conference.  Here I
discussed: electroweak tests; recent progress in the flavor sector; probes of
QCD in a variety of circumstances; and searches for new particles.  In the
third section, Mysteries, I returned to neutrinos to try to put into
context the implications of the SuperKamiokande results.  Finally, in the
Hopes section, I presented an outlook for the future, concentrating both
on windows of opportunities in the near term and the challenges posed by
new accelerators beyond the LHC.

\section{News}
\subsection{Atmospheric Neutrino Oscillations.}

The principal news of ICHEP 98 in Vancouver clearly was the evidence
presented by the SuperKamiokande collaboration for oscillations of atmospheric
neutrinos.~\cite{Takita,McGraw}  One has known for a number of years
that there are less $\nu_\mu$'s than expected coming from the atmosphere.\cite{R}  This trend was confirmed by the new SuperKamiokande
data, with the observed ratio of ratios for both sub-GeV $(E_{\rm vis} <
1.33~{\rm GeV})$ and multi-GeV $(E_{\rm vis} > 1.33~{\rm GeV})$ events
reported~\cite{McGraw}
\begin{equation}
R = \frac{\left(\frac{\nu_\mu}{\nu_e}\right)_{\rm data}}
{\left(\frac{\nu_\mu}{\nu_e}\right)_{\rm MC}} =
\left\{ \begin{array}{ll}
0.627^{\textstyle + 0.029}_{\textstyle - 0.027}
\pm 0.049 & {\rm (sub-GeV)} \\
0.647^{\textstyle + 0.052}_{\textstyle - 0.049}
\pm 0.08 & {\rm (multi-GeV)}
\end{array}
\right.
\end{equation}
being perfectly consistent with previous observations.  Although this
anomalous result is suggestive of neutrino oscillations, {\it per se} it
is not a convincing proof.

The observation by the SuperKamiokande collaboration of a clear zenith 
angle dependence of this signal provides strong evidence that neutrino
oscillations are indeed responsible for this phenomenon.  While the flux
of $\nu_e$ recorded is consistent with expectations, the $\nu_\mu$ flux
shows an anomalous zenith angle dependence.\cite{Takita,McGraw}
For multi-GeV events, since the geomagnetic field has little effect, one
expects both the $\nu_e$ and $\nu_\mu$ fluxes to be up-down symmetric.
For $\nu_\mu$ this expectation is belied by the SuperKamiokande data, as
shown in Fig. 1.  In an exposure of 535 days, 256 down-going multi-GeV
$\nu_\mu$ events were recorded, but only 139 up-going multi-GeV $\nu_\mu$ were
observed.

The SuperKamiokande collaboration~\cite{Takita,McGraw} interprets this
$6\sigma$ signal of an up-down $\nu_\mu$ asymmetry as evidence for 
$\nu_\mu\to \nu_X$ oscillations, with $\nu_X$ in this analysis being an
unspecified type of neutrino.  In terms of the usual two-flavor oscillation
formalism, typified by a mass squared difference $\Delta m^2$ and a mixing
angle $\theta$, the probability for a $\nu_\mu$ to survive is given by
\begin{equation}
P(\nu_\mu\to \nu_\mu) = 1-\sin^22\theta \sin^2
\left[\frac{1.27 \Delta m^2({\rm eV}^2) L({\rm Km})}{E_{\nu_\mu}
({\rm GeV})}\right]~.
\end{equation}
The SuperKamiokande $\nu_\mu$ data is consistent with nearly maximal mixing
$(\theta\simeq 45^o)$ and a mass squared difference
$\Delta m^2\simeq 10^{-3}~{\rm eV^2}$, with the best fit giving
$\sin^22\theta = 1$ and $\Delta m^2 = 2\times 10^{-3}~{\rm eV}^2$ with
a $\chi^2$ of 65.2 for 67 degrees of freedom.~\cite{McGraw}
\begin{figure}
\center
\epsfig{file=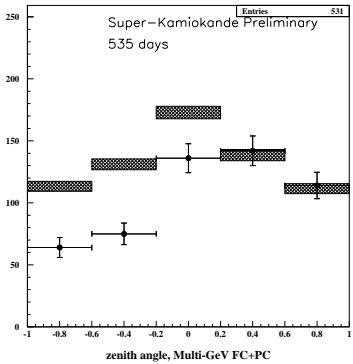,height=3in}
\caption{SuperKamiokande results on multi-GeV $\nu_{\mu}$ events.}
\end{figure}

At the moment, it is not possible to tell whether $\nu_X$ is a $\nu_\tau$
or possibly a sterile neutrino.  However, one can exclude the hypothesis
that one is dealing with a $\nu_\mu\to\nu_e$ oscillation.  The null
up-down $\nu_e$ zenith angle result from SuperKamiokande~\cite{Takita}
\begin{equation}
\left[\frac{U-D}{U+D}\right]_{\nu_e} =
0.036 \pm 0.067 \pm 0.020
\end{equation}
is $3\sigma$ away from what one would expect if a $\nu_\mu\to\nu_e$ oscillation
was involved.  Furthermore, the $\sin^22\theta-\Delta m^2$ region favored
by SuperKamiokande for $\nu_\mu\to\nu_e$ oscillations overlaps the region
{\bf excluded} by the CH00Z reactor experiment [$\langle E_\nu\rangle \sim
3~{\rm MeV};~L\sim 1~ {\rm Km}$] for $\nu_e\to\nu_\mu$ oscillations.~\cite{CH00Z}

I would like to make two comments on these results here, one experimental
and one theoretical.  On the experimental side, as Janet Conrad~\cite{Conrad}
nicely showed in her plenary talk, it is heartening to see that the SuperKamiokande
results are being corroborated by recent analyses of data from MACRO and
Soudan.  In addition, as Takita showed,\cite{Takita} there is internal 
consistency of the hypothesis of $\nu_\mu\to\nu_X$ oscillations in all the
different signals studied by both the Kamiokande and
SuperKamiokande experiments [contained
events; up-going through muons; and up-going stopped events].  The
apparent near non-overlap of the 90\% C.L. allowed region in the \break
$\Delta m^2-\sin^22\theta$ plane of the Kamiokande data with that from
SuperKamiokande, although at first sight perturbing, is an artifact of
focusing on the physical region.  The ``best fit" is actually unphysical,
with $\sin^22\theta|_{\rm K} = 1.35$ and $\sin^22\theta|_{\rm SK} = 1.05$,\cite{Conrad} but in the SuperKamiokande case this 
is only slightly preferred over the physical solution.
Thus, as Takita~\cite{Takita} emphasized in his plenary
talk, the data to focus on for its significance is that of SuperKamiokande.

Theoretically, these experimental results have important implications for
both particle physics and our understanding of the Universe.  First of all,
neutrino oscillations imply that neutrinos indeed have a mass.  Taking the
best fit value of SuperKamiokande for $\Delta m^2 = m_3^2-m_2^2 \simeq
2\times 10^{-3}~{\rm eV}^2$, one infers that at least one neutrino
(called $m_3$ here) is as heavy as $m_3 \stackrel{>}{_{\scriptstyle \sim}} 
5\times 10^{-2}~{\rm eV}$.
Such massive neutrinos contribute to the Universe's energy density at a
significant level, with the ratio of the neutrino density to the critical
closure density being more than one per mil:
\begin{equation}
\Omega_\nu = \frac{\rho_\nu}{\rho_c} \geq \frac{m_3~({\rm eV})}
{92~h^2} \sim 1.5\times 10^{-3}~.
\end{equation}
In the above $h$ is the scaled Hubble parameter, $h\sim 0.6$.  For
comparison, the contribution of luminous matter to the Universe's energy
density is of order $\Omega_{\rm lum} \sim (3-7)\times 10^{-3}~{\rm eV}$.
Thus, although neutrinos may not necessarily be the dominant component of
the dark matter in the Universe, they contribute the same as billions and
billions of stars!

For particle physics, the existence of a tiny neutrino mass
$(m_3\geq 5\times 10^{-2}~{\rm eV})$ is widely believed to be a 
natural reflection of a new
very heavy mass scale $M$, much heavier than the scale associated with the
breakdown of the electroweak theory $v \sim 250~{\rm GeV}$.  Although
$M \gg v$, the precise value for $M$ is somewhat model-dependent, but is of
order $10^{15}~{\rm GeV}$.  So, indeed, the SuperKamiokande results are
indications of ``new physics"!

Even though these arguments are 20 years old,\cite{seesaw} it may be helpful
to indicate briefly here why tiny neutrino masses naturally are tied to large
mass scales.  Because neutrinos are neutral, one has three possible neutrino
mass terms
\begin{equation}
{\cal{L}}_{\rm mass} = -\frac{1}{2} m_S\nu_{\rm R}^{\rm T}C \nu_{\rm R} -
m_{\rm D}\bar\nu_{\rm L}\nu_{\rm R} - \frac{1}{2} m_{\rm T}
\nu_{\rm L}^{\rm T}C \nu_{\rm L}~.
\end{equation}
Because $\nu_{\rm R}$ is an $SU(2)$ singlet and $\nu_{\rm L}$ is an
$SU(2)$ doublet, clearly the Dirac mass $m_{\rm D}$ and the Majorana mass
$m_{\rm T}$ must be proportional to the scale of the $SU(2)\times U(1)$
breakdown $v$.  The Majorana mass term involving $\nu_{\rm R}$ is
$SU(2)\times U(1)$ invariant and so $m_S$ is a new scale, independent of
$v$.

There are two cases to consider, depending on whether one assumes $\nu_{\rm R}$
exists or not.  In the latter case, then the observed neutrino mass is
$m_{\rm T}$.  If one assumes that the $SU(2)\times U(1)$ breaking is due
to a doublet Higgs, as there are good reasons to believe, then the triplet
mass $m_{\rm T}$ is proportional to $v^2$.  The formula
$m_{\rm T} \sim v^2/M_X$ numerically yields $M_X \sim 10^{15}~{\rm GeV}$
for $m_{\rm T}\sim 5\times 10^{-2}~{\rm eV}$.  If, on the other hand, 
$\nu_{\rm R}$ exists, then
the neutrino mass matrix is $2\times 2$.  Neglecting $m_{\rm T}$ and assuming
$m_S \gg m_D$,\cite{seesaw} the matrix
\begin{equation}
{\cal{M}} = \left(
\begin{array}{cc}
m_{\rm T} & m_{\rm D} \\
m_{\rm D} & m_{\rm S} 
\end{array} \right) \simeq
\left(
\begin{array}{cc}
0 & m_{\rm D} \\
m_{\rm D} & m_{\rm S}
\end{array} \right)
\end{equation}
has both a light state, of mass $m^2_{\rm D}/m_{\rm S}$, and a superheavy state
of mass $m_{\rm S}$.  If one identifies $m_3$ 
as the $\nu_{\tau}$ mass, it is not unnatural to equate the neutrino Dirac mass $m_{\rm D}$ with the top mass.  The formula
$m_3 \simeq m_t^2/m_S$ again suggests a mass scale $m_S \sim 10^{15}~{\rm GeV}$.

Although the above simple discussion can at best only be indicative of the
order of magnitude of the mass scale associated with the presence of sub eV neutrino masses, I find it remarkable that the scale that emerges is of the order of the
GUT scale $[M_{\rm GUT} \sim 2\times 10^{16}~{\rm GeV}$, 
obtained from gauge coupling
unification (with supersymmetry).\cite{unification}]  In a way 
this result represents
a vindication of history, since the SuperKamiokande experiment is a direct
descendant of the original experiment in the Kamioka mine looking for
proton decay.\cite{Kamioka}  Although there is still no proton decay signal,
with SuperKamiokande pushing the limit for the
$p\to\nu K^+$ mode to $\tau/B(p\to\nu K^+) > 5.5\times 10^{33}$ years,~\cite{Hayato} perhaps indirectly through neutrino oscillations we are
probing this same physics---a point emphasized by Babu~\cite{Babu} at this
Conference.

\subsection{Recent Results on Cosmological Parameters.}

In the past year there has been a sharpening and shift in the value of the
cosmological parameters,\cite{Spiro,Pain} with interesting
implications both for high energy physics and for cosmology.\cite{Kolb}
The ``Standard Picture" of cosmology assumes that $\Omega = \rho/\rho_c$ is
unity, as predicted by inflation---corresponding to a flat Universe.  Pre
1998, it was believed that dark matter was the dominant component of $\Omega$
($\Omega_{\rm DM} \simeq 0.95$, with perhaps $\Omega_\nu \sim 0.2$~\cite{Primack})
with a small amount of the Universe's energy density in the form of baryons
($\Omega_B \simeq 0.02-0.08$, from primordial nucleosynthesis~\cite{ns}) and
no cosmological constant contribution $(\Omega_\Lambda = 0)$.  A variety of
observations in the last year have changed this state of affairs 
considerably.  If $\Omega = 1$, as inflation suggests, then now it appears
that
\begin{equation}
\Omega_M = \Omega_{\rm B} + \Omega_{\rm DM} \simeq 0.35~; ~~~~
\Omega_\Lambda \simeq 0.65~.
\end{equation}

These results emerge from rather different sets of observations.\cite{Kolb}
I indicate here some of them, to give a flavor of their nature:
\begin{description}
\item{(i)} By studying the nature of galaxy clustering at different
scales one has now considerably sharpened the estimate of $\Omega_M$.
Coupled with a now rather precise determination of the Hubble parameter
$(h = 0.6\pm 0.1)$,~\cite{Pain} this considerably narrows the range for
$\Omega_M$ allowed by earlier estimates,~\cite{DBW} with  values of $\Omega_M \sim 0.3$ favored. ~\cite{Kolb}
\item{(ii)} If $\Omega_M$ were to be near unity, one would expect very few
large galaxy clusters at high red shift. ~\cite{cohn}  Recent observations,~\cite{HB}
in fact, see an order of magnitude more large clusters 
than expected, consistent
with $\Omega_M \simeq 0.3$ in a flat Universe.\cite{VL}
\item{(iii)} By measuring Type Ia supernovas at very large red shift,~\cite{Pain,Spiro} a measurement of the deceleration parameters 
$q_0$ for the Universe is possible.  Since
\begin{equation}
q_0 = \frac{\Omega_M}{2} - \Omega_\Lambda~,
\end{equation}
this parameter measures a different combination of $\Omega_M$ and $\Omega_\Lambda$  than the total energy density. 
Thus, in principle, one can hope to
disentangle the two.  The expectation for a matter dominated flat
Universe is that $q_0 \simeq 1/2$.  What was seen instead was that
$q_0 < 0$, corresponding to an accelerated expansion.  Fig. 2 displays the
results of the Supernova Cosmology Project~\cite{SCP} presented at this
meeting by Pain.\cite{Pain}  Clearly an $\Lambda=0$ flat Universe is excluded
at the many standard deviation level. 
\end{description}
\begin{figure}
\center
\epsfig{file=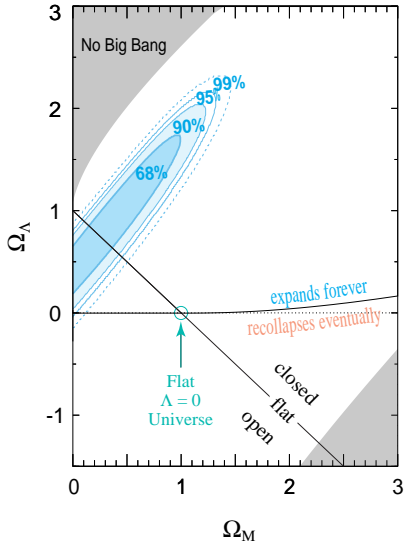,height=3in}
\caption{Supernova Cosmology Project results.}
\end{figure}

Although these new results from cosmology are very interesting, they need
further checking.  For example, there could well be unknown systematics
in the Supernova data.  Fprtunately,
a very sensitive check of the cosmological parameters 
will be provided by the MAP and
Planck satellite experiments presently under construction, since a precise
measurement of the cosmic microwave angular spectrum is a very sensitive probe
of the various density components.  

If the 
present results hold true, they actually
have a potentially significant impact for particle physics.  For instance,
if $\Omega_{\rm DM} \simeq 0.3$ this may actually help rather than hinder
experiments looking for WIMPs.\cite{Bauer}  The density of WIMPs in the galaxy
should be unaffected, since it is a function of galactic parameters.
However, since in general $\sigma_{\rm WIMP} \sim 1/\Omega_{\rm DM}$, the
event rate will go up rather than down!

Once one admits that $\Omega_\Lambda \not= 0$, then there is less need of
having a hot dark matter component to fit the spectrum of galaxy 
fluctuations.\cite{Primack}  At any rate, if $\Omega_{\rm DM} \simeq 0.3$,
certainly $\Omega_\nu$ is at most of order of $\Omega_B$, not around 0.2.
Thus, the likely range of neutrino masses which are cosmologically significant
is reduced to perhaps
\begin{equation}
\sum_i m_{\nu_i} \sim (0.5-2)~{\rm eV}~,
\end{equation}
an unfortunate range for short baseline experiments!

Finally, even though $\Omega_\Lambda$ may well be non-zero, it seems clear
to me that whatever its source, it is not a signal of a particle physics
``vacuum energy".  Efforts should persist to try to understand why the
cosmological constant {\bf vanishes} theoretically, since it is certainly
easier to try to explain zero than the value one infers from $\Omega_\Lambda$
\begin{equation}
\langle T^\mu_\mu \rangle_\Lambda \simeq
(3\times 10^{-3}~{\rm eV})^4~,
\end{equation}
which involves an absurd particle physics scale!

\section{Refinements}
\subsection{All is Well in the Electroweak Front.}

Substantial new data has sharpened our understanding of the electroweak
theory and its parameters.  The net result, as Wolfgang Hollik~\cite{Hollik}
and Dean Karlen~\cite{Karlen} made clear in their plenary talks, is that the
electroweak theory is totally consistent with the precision electroweak 
data at the 0.1\% level---a remarkable fact.  Let me briefly review some of the
main points.

\subsubsection{Top.}

The CDF and D0 combined results for the top mass, presented at this
conference by E. Barbieris,\cite{Barbieris} now make top the quark whose
mass is best known.  The combined result
\begin{equation}
m_t = (173.8 \pm 3.2 \pm 3.9)~{\rm GeV} =
(173.8 \pm 5.0)~{\rm GeV}
\end{equation}
has a relative error $\delta m_t/m_t$ of less than 3\%---a remarkable result.
In addition, CDF and D0 have a quite accurate determination of the top pair
production cross section~\cite{Partridge}
\begin{equation}
\sigma_{t\bar t} =
\left\{ \begin{array}{ll}
(5.9 \pm 1.7)~{\rm pb} & {\rm D0} \\
\left(7.6^{\textstyle + 1.8}_{\textstyle -1.5}\right)~{\rm pb}
& {\rm CDF}
\end{array}
\right.
\end{equation}
in good agreement with the theoretical QCD predictions, which range from
4.7 to 6.2 pb.

It was apparent in this Conference that the study of top
at the Tevatron is entering a more mature phase, moving from a period of
discovery to one where one is trying to characterize top's properties.  For
instance, in the Standard Model one expects that in the dominant top decay,
$t\to Wb$, the produced $W$ is longitudinally polarized about 70\% of the time.
This prediction is borne out by data from CDF, discussed by Tollefson,~\cite{Tollefson}
which determined the fraction of longitudinally
polarized $W$'s produced in top decay to be
\begin{equation}
F_{\rm L} = 0.55 \pm 0.32 \pm 0.12~.
\end{equation}

\subsubsection{$W^\pm$.}

Information on the $W$ mass came from a number of different quarters at
Vancouver.  Indirectly, the NUTEV experiment at Fermilab was able to infer
a rather precise value for $M_W$ from a high statistics neutrino deep inelastic
experiment.\cite{NUTEV}  By using sign selected beams, the NUTEV
collaboration was able to largely avoid the uncertainty caused by the
charm component in the nucleon in its measurement of the weak angle.
Specifically, NUTEV effectively measured the Paschos-Wolfenstein 
ratio~\cite{PW}
\begin{equation}
R = \frac{\sigma^\nu_{\rm NC} - \sigma^{\bar\nu}_{\rm NC}}
{\sigma^\nu_{CC} - \sigma^{\bar\nu}_{\rm NC}} = \frac{1}{2} -
(\sin^2\theta_W)_S~,
\end{equation}
thereby reducing the error on $\sin^2\theta_W$ due to the unknown charm
contribution by at least 50\% compared to the error reported by the 
CCFR collaboration.
The new NUTEV result, presented by T. Bolton,\cite{NUTEV} gives a mixing angle
\begin{equation}
(\sin^2\theta_W)_S \equiv 1-M^2_W/M^2_Z =
0.2253 \pm 0.0019 \pm 0.0010~,
\end{equation}
which determines the $W$-mass to an accuracy of 110 MeV~\cite{NUTEV}
\begin{equation}
M_W = (80.26 \pm 0.11)~{\rm GeV}~.
\end{equation}

This result is in agreement with the more precise values for $M_W$ inferred
from studies of the process $e^+e^-\to W^+W^-$ at LEP2 and from $W$
production at the Tevatron.  
Combining the threshold analysis of the $W$ mass at
$\sqrt{s} = 161$ GeV with the value of $M_W$ obtained by
direct reconstruction at both $\sqrt{s} = 172$ GeV and $\sqrt{s} = 183$ GeV,
the averaged results from the four LEP collaborations determine the $W$ mass
to 90 MeV~\cite{Thompson}
\begin{eqnarray}
M_W &=& (80.37 \pm 0.07 \pm 0.04 \pm 0.02)~{\rm GeV} \nonumber \\ 
&=& (80.37 \pm 0.09)~{\rm GeV}~.
\end{eqnarray}
In the above, the dominant error (70 MeV) is statistical, with about
40 MeV coming from not being able to disentangle final state interactions
between the two produced $W$'s and 20 MeV arising from uncertainties in the
beam energy.

A similar error is obtained by combining the values for $M_W$ obtained by the
CDF and D0 collaborations (as well as the old UA2 data),~\cite{Kotwal}
giving a ``collider value" for $M_W$ of
\begin{equation}
M_W = (80.40 \pm 0.09)~{\rm GeV}~.
\end{equation}
The World average for $M_W$, determined from the above 2 direct measurements
has an error of 60 MeV~\cite{Karlen}
\begin{equation}
\left. M_W\right|_{\rm direct} = (80.39 \pm 0.06)~{\rm GeV}~,
\end{equation}
so that now we know the $W$ mass to better than 1 part per mil.
This value is in excellent agreement not only with the NUTEV result, but
also with the very precise indirect $W$ mass determination obtained from a
global fit of all other high precision electroweak data, which
gives~\cite{Karlen}
\begin{equation}
\left. M_W\right|_{\rm indirect} = (80.365 \pm 0.030)~{\rm GeV}~.
\end{equation}
I comment below on this latter fit and its implications.

\subsubsection{Standard Model Tests.}

Precision measurements at the $Z$ resonance, plus a knowledge of $m_t$
and $M_W$, overconstrain the Standard Model.  Thus, as Karlen~\cite{Karlen}
and Hollik~\cite{Hollik} emphasized, present-day data provides rather
significant tests of the electroweak theory.  Fits of all 
electroweak data to the
Standard Model are in terrific agreement with expectations, with very few quantities in the fit
being over $2\sigma$ away from the fit value.  This is nicely seen in
Fig. 3, which summarizes the Standard Model analysis presented by 
Gr\"unewald~\cite{G} in Vancouver.  Not only is the data consistent with the
Standard Model, but it is also internally consistent.  This was most
clearly seen in the comparison of different determinations of $\sin^2\theta_W$
at both LEP and SLD, presented by Baird,\cite{Baird} which also lay at most
$2\sigma$ away from the average value
\begin{equation}
\sin^2\theta_W^{\rm eff} = 0.23155 \pm 0.00018~.
\end{equation}

\begin{figure}
\center
\epsfig{file=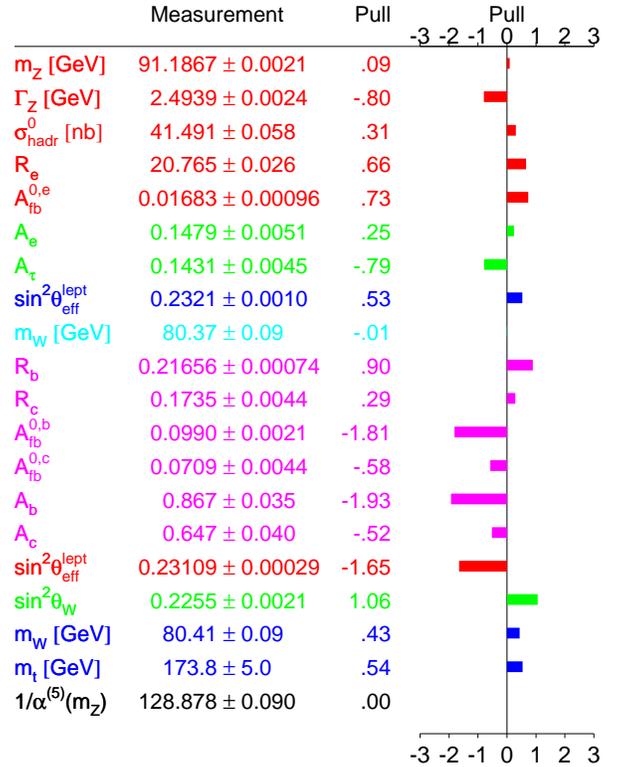,height=4in}
\caption{Standard Model Fit.}
\end{figure}

This amazing experimental precision is being matched theoretically.  As
J. H. K\"uhn~\cite{K} discussed in his talk in the parallel sessions, at present
both 2-loop electroweak, and mixed QCD-electroweak, corrections are being
incorporated in the fitting programs.  Typically, these corrections
contribute to the $W$ mass at the level of 10 MeV, to be contrasted to the
30 MeV error of the global fit.

In the Standard Model, given $G_F,~\alpha,~M_Z$ and $m_t$, the only free
parameter remaining is the Higgs mass $M_H$.\cite{Hollik}  Unfortunately,
even the present high
precision data does not give a strong constraint on $M_H$, since the
effects of the Higgs mass are only proportional to $\alpha~\ell n~M_H$.
Nevertheless, the 68\% CL contours in the $M_W-m_t$ plane shown by Karlen,\cite{Karlen} determined both through the Standard Model fits and by
the direct measurements of $m_t$ and $M_{\rm W}$, favor a low value for the Higgs
mass
\begin{equation}
M_H = \left(84^{\textstyle + 91}_{\textstyle -51}\right)~{\rm GeV}~,
\end{equation}
leading to a one-sided 95\% CL bound for $M_H$ of $M_H < 280~{\rm GeV}$.

\subsubsection{Direct Higgs Searches.}

At the Conference, improved lower bounds on the Higgs mass coming from
LEP2 were also reported, from searches of the process $e^+e^-\to ZH$.  The
combined limits from all four
LEP experiments, using the $\sqrt{s} = 183~{\rm GeV}$
data~\cite{McNamara,Treille} give at 95\% C.L.
\begin{equation}
M_H > 89.8~{\rm GeV}
\end{equation}
A ``first look" of about $35~{\rm pb}^{-1}$/experiment at
$\sqrt{s} = 189~{\rm GeV}$ already gives similar results.\cite{McNamara}
Indeed, the preliminary data from OPAL, which has less background events
than expected, gives a stronger limit: $M_H > 93.6~{\rm GeV}$.
Nevertheless, more definite  results from $\sqrt{s} = 189~{\rm GeV}$ must
await the end of the ongoing run.

\subsection{Sharpening the CKM Parameters.}

Considerable progress was also reported at ICHEP 98 in the flavor sector,
with various important branching ratios sharpened and some of the CKM matrix
elements determined with greater precision.~\cite{Bryman,Alexander}
Here I focus on some of the most notable results.

\subsubsection{$K\to\pi\nu\bar\nu$.}

The observation by the Brookhaven experiment E787 of one event for the charged
kaon decay mode $K^+\to\pi^+\nu\bar\nu$, giving a branching ratio~\cite{Red}
\begin{equation} 
{\rm BR}(K^+\to\pi^+\nu\bar\nu) = 
\left(4.2^{\textstyle
+ 9.7}_{\textstyle -3.5}\right) \times 10^{-10}~,
\end{equation}
allows one to infer a range for $V_{\rm td}$: $0.006 \leq |V_{\rm td}|
\leq 0.06$ directly from the Kaon sector.  The KTeV collaboration~\cite{KTeV}
also reported a new 90\% C.L. bound on the process $K^0_{\rm L}\to
\pi^o\nu\bar\nu$
\begin{equation}
{\rm BR}(K_{\rm L}^o \to \pi^o \nu\bar\nu) < 5.9 \times 10^{-7}~.
\end{equation}
This result constrains the CP-violating parameter $\eta$, in the Wolfenstein
parametrization ~\cite{Wolfenstein} of the CKM matrix, to $\eta < 50$.
Although both the $|V_{\rm td}|$ determination and this bound are less 
precise than extant values, they highlight the continuing potential of
rare $K$ decays for our understanding of the CKM matrix.

\subsubsection{T-Violation.}

Because of the CPT theorem,\cite{CPT} all measurements of CP-violation are
also a measure of T-violation.  However, up to now, no direct measurement
of a T-violating asymmetry had been reported.  This was remedied at
Vancouver, where the CP Lear collaboration reported the first measurement of the
T-violating Kabir asymmetry~\cite{Kabir}
\begin{equation}
A_{\rm T}(t) = \frac{\Gamma(\bar K^0 \to \pi^-e^+\nu(t)) -
\Gamma(K^0\to\pi^+e^-\bar\nu(t))}{\Gamma(\bar K^0\to\pi^-e^+\nu(t)) +
\Gamma(K^0\to\pi^+e^-\bar\nu(t))}~.
\end{equation}
For large times, $t\gg\tau_s$, $A_{\rm T}$ measures (assuming CPT holds)
the well known CP violating parameter in the Kaon system
4 Re $\epsilon$.
The CP Lear result~\cite{Kokkas}
\begin{equation}
A_{\rm T}(t \gg \tau_s) = (8.0 \pm 1.7 \pm 1.0) \times 10^{-3}
\end{equation}
is consistent, within errors, with this expectation.

\subsubsection{B-Decays.}

Alexander~\cite{Alexander} discussed a refined measurement by CLEO (and the
first result of ALEPH~\cite{A}) of the branching ratio for the process
$B\to X_s\gamma$.  The CLEO result
\begin{equation}
{\rm BR}(B\to X_s\gamma) =
(3.15 \pm 0.35 \pm 0.32 \pm 0.26)\times 10^{-4}
\end{equation}
(where the last error is an estimate of the model dependence uncertainty) is
in excellent agreement with the expectations of the Standard Model,
including non-leading order QCD corrections, reported by Neubert~\cite{Neubert}
in the parallel sessions
\begin{equation}
{\rm BR}(B\to X_s\gamma)|_{\rm SM} =
(3.29 \pm 0.33) \times 10^{-4}~.
\end{equation}
A comparison of theory with experiment~\cite{Alexander} allows one to infer
a value for $|V_{ts}|$ with an error of around $10\%$ - $|V_{ts}| = 0.035
\pm 0.004$.  Barring cancellations, these results also give a rather
tight bound on the mass of a hypothetical charged Higgs:
$M_{H^+} > 210~{\rm GeV}$. Including QED and electroweak corrections, this bound goes down to $M_{H^+}>165$ GeV. ~\cite{greub}

Study of $B_s-\bar B_s$ oscillations in $Z^0$ decays at LEP and the SLD,
reported by Parodi,~\cite{Parodi} provide a new lower bound for
$\Delta m_s$:
\begin{equation}
\Delta m_s > 12.4~{\rm ps}^{-1}~.
\end{equation}
This result, combined with new refined measurements of $V_{ub}$ both in the
exclusive mode $B\to \rho\ell\nu$~\cite{Alexander} at CLEO and from
inclusive $B\to X_u\ell\nu$ studies at CLEO and LEP,\cite{Rosnet} giving
\begin{equation}
|V_{ub}| = (3.56 \pm 0.21 \pm 0.28 \pm 0.43)\times 10^{-3}
\end{equation}
(with again the last error being a ``theory" error), allow one to further
restrict the allowed region in the $\rho-\eta$ plane for the CKM model.
Fig. 4 displays the results of this analysis, presented by Parodi~\cite{Parodi}
in Vancouver.

\begin{figure}
\center
\epsfig{file=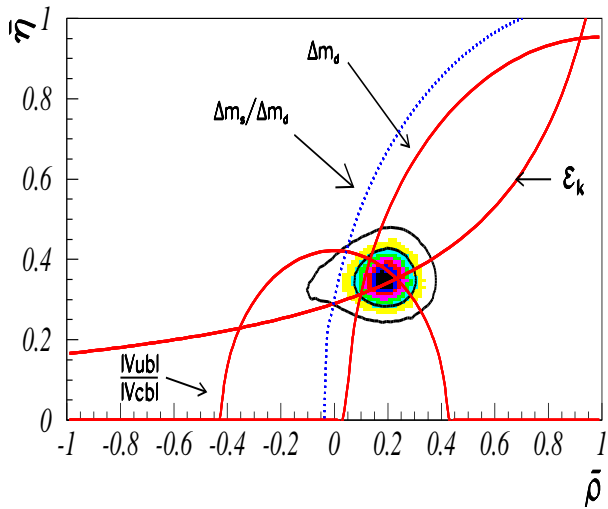,height=3in,width=4in}
\caption{Allowed region in $\rho-\eta$ plane.}
\end{figure}

\subsection{QCD at Work.}

In contrast to the electroweak and flavor sector, QCD is a rather mature
theory.  Thus, as Yuri Dokshitzer~\cite{Dok} emphasized in his plenary talk,
the issue is not really ``to check QCD, but rather understand how it
works".  Theoretically, this understanding comes from two fronts:
lattice QCD~\cite{Sharpe} and from applications of perturbative QCD~\cite{Dok}
in circumstances where one has some control.

\subsubsection{Lattice Results.}

Steve Sharpe~\cite{Sharpe} in his plenary talk reported significant progress in two areas from
lattice QCD calculations.
\begin{description}
\item{(i)} The CP-PACS collaboration~\cite{CPPACS} obtained rather accurate
results for the hadronic spectrum, in the quenched approximation, but close
to the chiral limit and with rather fine lattices (a = 0.05 fm).  Typically,
these results reproduce the known spectrum of hadrons to within 10-20\%,
with the errors argueably arising from the neglect of quark loops.
\item{(ii)} A variety of weak matrix elements, calculated in the quenched
approximation, but with estimates of the possible unquenched contributions, 
also
have similar errors.  For instance, the parameter $\xi$ needed to extract
$|V_{td}|/|V_{ts}|$ from the comparison of $B_d-\bar B_d$ mixing to
$B_s-\bar B_s$ mixing emerging from the best lattice calculation,\cite{Draper}
\begin{equation}
\xi = \frac{f_{B_s}\sqrt{\hat B_{B_s}}}
{f_B\sqrt{\hat B_B}} = 1.14 \pm 0.06 \pm 0.03 \pm 0.10~,
\end{equation}
has an 0.06 error from the calculation itself, with an estimated error of
0.03 from the unquenched contribution and 0.10 from chiral loops.\cite{Sharpe}
\end{description}

\subsubsection{Perturbative QCD Results.}

Deep inelastic scattering provides the theoretically most pristine arena
for seeing QCD at work.  This was apparent in the beautiful HERA data on
$F_2(x,Q^2)$ shown by Doyle,\cite{Doyle} where one could see with the
naked eye very little  
$Q^2$-dependence at large $x$, but significant $Q^2$-dependence
at low $x$.  Although not a test of QCD, it was also nice to see that at
large $Q^2$ CC scattering and NC scattering at HERA are comparable, as
expected.  

Doyle~\cite{Doyle} in his plenary talk discussed also other data which is quite important
for QCD, connected with the spin structure functions.  Here all experimental
results, including the latest E155 data from SLAC, are converging and they
give a consistent picture for the spin sum rules, irrespective of whether
one is scattering polarized leptons off protons, neutrons, or deuterons.
Furthermore, the SMC Collaboration global NLO QCD analysis of their
data for the Bjorken sum rule, reported by Doyle,\cite{Doyle}
\begin{equation}
\int^1_0 dx (g_1^p(x;Q^2)\left. - g_1^n(x;Q^2))\right|_{Q^2=5~{\rm GeV}^2}
= 0.175^{\textstyle + 0.024}_{\textstyle -0.012}
\end{equation}
agrees to 10\% with the theoretical predictions calculated to
$O(\alpha_s^3)$ [Theory: $0.181 \pm 0.003$].  So everything seems to be in
order in deep inelastic scattering, at least in areas where one can
calculate reliably.

This is also the case for the study of jets at large momentum transfer.  Here
the inclusive jet distributions, irrespective of whether the jets are
observed at HERA~\cite{Sinclair} or at the Tevatron,~\cite{Blazey} are
extremely well fit by QCD over many decades.\cite{Huston}
The ``anomaly" at very large $E_T$ reported by CDF~\cite{largeE} has not
dissapeared, but is not really seen in the D0 data.  In fact, the CDF and D0
data are quite consistent,\cite{Huston} with the only extant discrepancy being
a 20\% discrepancy between theory and experiment in the ratio of large
$E_T$ data at $\sqrt{s} = 630~{\rm GEV}$ to data at $\sqrt{s} = 1800~{\rm GeV}$.

The comparison of data with QCD is more complicated for more differential
observables.  Nevertheless, as Dokshitzer~\cite{Dok} remarked, because
confinement is rather ``soft" no large color fields appear.  This bolsters
the hope that one can actually apply perturbation theory down to very low
$Q^2$---perhaps as low as $Q^2\sim 2~{\rm GeV}^2$---and in a variety of
circumstances.

The results presented at Vancouver on event shapes and fragmentation 
studies at LEP, as well as dijet production at both HERA and the Tevatron,
seem to underline this point.  For instance, Duchesnau~\cite{Duchesnau}
presented LEP data which clearly showed evidence of gluon coherence, with
the peak of the spectrum of charged particles plotted versus $\xi =
\ln 1/x$ clearly growing with $\ln E_{\rm beam}$.  At HERA the study of dijet
production~\cite{Huston} and comparison with perturbative QCD predictions
allows a preliminary extraction of the gluon distribution function at small
$x$, showing its expected growth as $x\to 0$.  One can also extract directly
$\alpha_S(M_Z)$ from studying various event shape variables~\cite{Duchesnau}
at LEP.  The resulting value of $\alpha_S(M_Z)$ obtained is in good agreement
with other determinations, summarized by Bethke,~\cite{Bethke} which lead to
a world average value
\begin{equation}
\alpha_S(M_Z) = 0.1190 \pm 0.0058
\end{equation}

\subsection{Searches for New Physics.}

Many beautiful searches for new phenomena were presented at Vancouver, but
no clear signal was found.  This broad subject matter was expertly
summarized by Daniel Treille~\cite{Treille} who pointed out how the
various results presented have served to sharpen the bounds on hypothetical
particles.  Often these new bounds arise in a complementary fashion from
different machines.  A good case in point are bounds on first generation
leptoquarks.  Here HERA's bounds depend both on the strength of the
leptoquark coupling and its mass, while the bounds coming from the Tevatron
depend only on the leptoquark mass.  For couplings of electromagnetic
strength, these bounds are quite comparable.\cite{Treille}

Most of the attention now is concentrated on searches for supersymmetric
partners of the quarks, leptons and gauge fields we know.  These searches,
however, are both not easy experimentally and not simple to describe in
detail.  The results one gets depend intrinsically on what assumptions
one makes on how exactly supersymmetry is broken, since these assumptions 
change the nature of the lightest supersymmetric particles (LSP) expected.
In supergravity theories, where the breaking of supersymmetry occurs in a
hidden sector connected to the observable sector by gravitational strength
interactions, the LSP is a neutralino.  On the other hand, in theories where
the breaking of supersymmetry is mediated by gauge interactions, the
LSP is always the gravitino.  However, in this case the important
excitation to focus on is the next lightest supersymmetric particle,
the NLSP.  Furthermore, the searches for supersymmetric partners are
different depending on whether or not $R$-parity is conserved or broken.
In the latter case, the LSP is actually unstable and can decay into
ordinary particles, complicating the searches further.

For these reasons, it is not possible to state model independent bounds on
supersymmetric partners, without some associated theoretical framework
underpinning these bounds.  This is somewhat easier for Higgs searches,
since in supersymmetric theories, one always needs to have at least two
different Higgs doublets.  Hence, in this case, 
one expects instead of the single Standard
Model Higgs boson, the neutral states ($h,H$ and $A$) and a pair of charged
states $(H^\pm)$.  The searches at LEP2 for the Standard Model Higgs also provide
strong bounds for the lightest scalar $(h)$ and pseudoscalar $(A)$ Higgs
bosons.  Combining the results of all four LEP collaborations one finds~\cite{Desch}
\begin{eqnarray}
m_h &>& 77~{\rm GeV} \\
m_A &>& 78~{\rm GeV}
\end{eqnarray}

\section{Mysteries}

There are many mysteries in our field, ranging from the origin of the
$SU(2)\times U(1)$ breakdown, to the reason for families and the peculiar
spectrum of quark and leptons.~\cite{Nelson}  In Vancouver, these mysteries,
however, took a back seat to what I called the neutrino maelstr\"om.

\subsection{The Neutrino Maelstr\"om.}

There are actually two different aspects of the neutrino maelstr\"om.  The
first concerns what data one actually believes concerning neutrino masses,
mixing and oscillations.~\cite{Takita,Conrad}  The second is
connected with what theoretical prejudices guide one's thinking.\cite{Kayser,Nandi,Babu}  I want to convey some flavor
of the controversies connected with these two points, since it is precisely
these kinds of controversies which help to keep our field healthy and alive.

Besides the SuperKamiokande evidence for atmospheric neutrino oscillations
there are two other oscillations hints, the solar neutrino deficit and the
LSND signal.  In addition, there are two different bounds on the mass of the
neutrino coupled preferentially to the electron.  Let me briefly touch on
these latter bounds first.

\subsubsection{Bounds on $m_{\nu_e}$.}

Tritium beta decay experiments are sensitive down to a ``few eV" for
$\nu_e$ masses.  However, all the highest precision experiments actually see
an excess of events near the end point, leading to a tachyonic neutrino
mass~\cite{PDG}
\begin{equation}
\langle m^2_{\nu_e}\rangle = (-27 \pm 20)~{\rm eV}^2~.
\end{equation}
Because these excess events are not understood, it is not possible to quote
a bound on $m_{\nu_e}$ at the level of sensitivity of these experiments.

Double beta decay experiments, principally using ${}^{76}{\rm Ge}$, also
provide a strong bound to the combination of neutrino masses which couple to
the electron.  The bounds one obtains~\cite{Kayser} have some range, due to
uncertainties in the corresponding nuclear matrix elements, but are also at the
eV level:
\begin{equation}
\langle m_\nu\rangle_{ee} = \sum U^2_{ei} m_{\nu_i} \leq
(0.5-1.5)~{\rm eV}~.
\end{equation}
In the above $U_{ei}$ is the mixing matrix element connecting the
$\nu_e$ neutrino to the i$^{th}$ mass eigenstate.  This is an important
bound if neutrinos are Majorana particles, but the bound evaporates if
neutrinos are Dirac particles, since double beta decay cannot proceed
if fermion number is conserved.

\subsubsection{The Solar Neutrino Deficit.}

At present, all five experiments [Homestate, SAGE,
Gallex, Kamiokande, and
SuperKamiokande] which measure solar neutrinos appear to
record~\cite{Conrad} roughly 50\% of the rate expected by the Standard Solar
Model (SSM).\cite{SSM}  This is the case, 
notwithstanding the fact that these
experiments are sensitive to different components of the solar neutrino
spectrum.  For instance, both SAGE and Gallex are mostly sensitive to 
neutrinos arising in the $pp$ cycle, while both Kamiokande and
SuperKamiokande essentially are only sensitive to ${}^8B$ neutrinos.

The observed experimental deficit has two possible interpretations, within
the context of neutrino oscillations.\cite{Conrad}  There is a ``just so"
solution, in which the results seen are interpreted as $\nu_e\to \nu_X$
oscillations with maximal mixing and with $\Delta m^2\sim 10^{-10}~{\rm eV}^2$.
With such a small mass squared difference, given that $E_\nu\sim O({\rm MeV})$
and $L \sim 10^8~{\rm Km}$, the oscillating factor $\sin^2(1.27 \Delta m^2
L/E)$ in the transition probability averages to 1/2.  However, the
fact that any solar neutrino deficit is observed is due to the good fortune 
that the sun is precisely far
enough away from earth to make the effect visible---a rather lucky
coincidence!

A more likely interpretation, instead, is that the reduced flux seen is the
result of matter induced oscillations,\cite{MSW} in which $\nu_e$'s produced
in the sun's interior totally convert to some other neutrino species as
they transverse regions with different electron density.  All experimental
data are consistent with having $\Delta m^2 \sim 10^{-5}~{\rm eV}^2$ and
either $\sin^22\theta\sim 3\times 10^{-3}$ (small angle solution) or
$\sin^22\theta\sim O(1)$ (large angle solution).\cite{Hata}

Of course, either the ``just so" or the matter induced oscillation
interpretation of the solar neutrino data relies on the validity of the
Standard Solar Model.  An issue often raised is how reliable this model is.
In particular, one knows that the detailed flux of ${}^8B$ neutrinos is rather
sensitively dependent on the temperature of the solar core [$\Phi({}^8B~\nu_e)
\sim T^n_{\rm core}$ with $n\sim (18-24)$~\cite{Kayser}].  
Fortunately, recent results 
from helioseismology appear to be in excellent agreement with the core
temperatures predicted by the SSM, bolstering the arguments for the
oscillation iterpretation of the solar neutrino data.

\subsubsection{The LSND Signal.}

The LSND collaboration~\cite{LSND} summarized in Vancouver their evidence for
neutrino oscillations, in which either a $\bar\nu_\mu$ or a $\nu_\mu$ appear
to transmute themselves into a $\bar\nu_e$ or $\nu_e$,
respectively.  These observed signals are consistent with a mass squared
difference of $O(\Delta m^2\sim 0.1~{\rm eV}^2)$ and rather small mixing
angles.\cite{Conrad}  The LSND signal, however, is somewhat suspect because
it lays very close to the region in $\Delta m^2-\sin^22\theta$ excluded by other experiments already.  To avoid these already excluded regions~\cite{Conrad}
one must assume that $\sin^22\theta \stackrel{<}{_{\scriptstyle \sim}} 2\times 10^{-2}$
and $\Delta m^2 < {\rm eV}^2$.

Even if $\Delta m^2$ and $\sin^22\theta$ obey the above limits, data from 
Karmen presented at this conference~\cite{Karmen} appears to contradict the
LSND results.  However, as Conrad emphasized,\cite{Conrad} one has to be
quite careful in not making too strong a statement here.  In their running to date,
Karmen sees no events indicative of $\nu_\mu\to\nu_e$ oscillations, expecting
about 3 events of background and 1 event of signal (taking LSND at face
value).  Not seeing any events allows Karmen to exclude the LSND
signal at 90\% confidence.  However, in fact, if they had observed the
background events expected, the Karmen sensitivity would not 
have been enough by itself
to exclude the LSND result to 90\% confidence.  Hence, it may prudent to
wait for more data before making a definitive pronouncement!

\subsection{Theoretical Considerations.}

The theoretical scenarios pursued to explain the hints for neutrino masses
and mixings reflect more the prejudices of the practioners rather than
some deep-seated truths.  Roughly speaking, these scenarios break up into 
two different classes, depending on whether one believes or not all data.
If one takes all data hints for neutrino masses at face value then the most
natural scenarios requires the introduction of a sterile neutrino $\nu_S$,
which by its nature does not couple to the $Z$ boson.

A typical model including a sterile neutrino is that of Caldwell and Mohapatra.~\cite{Caldwell}  In their case, one interprets the solar neutrino
deficit as resulting from a $\nu_e\to\nu_S$ oscillation, with mass squared
differences of order $\Delta m^2 \simeq 10^{-5}~{\rm eV}^2$.  Atmospheric neutrino 
oscillations are attributed to $\nu_\mu\to\nu_\tau$ oscillations with
$\Delta m^2_{\rm atmos}\sim 10^{-3}~{\rm eV}^2$ and the LSND signal is due to
$\nu_\mu\to\nu_e$ oscillations with $\Delta m^2_{\rm LSND}\sim 10^{-1}~{\rm eV}^2$.
Because one has now four different neutrinos, these schemes naturally can
introduce three different $\Delta m^2$ values.

As Conrad~\cite{Conrad} pointed out in her talk, however, it is also possible
to fit all three oscillation signals {\bf without} introducing a sterile
neutrino.  To be able to do so, it is necessary to assume that the atmospheric
neutrino signal involves simultaneously $\nu_\mu\to\nu_e$ and
$\nu_\mu\to\nu_\tau$ oscillations.  The resulting models ~\cite{stretch} require
considerable data stretching and, in my view, are marginally viable.

Personally, I believe that it is much more likely that {\bf not} all of the
hints for neutrino oscillations are true.  Eliminating one of the oscillation
hints allows one to deal only with the known neutrinos and attribute each of
the observed oscillation signals mainly to one given oscillation mode.
Nevertheless, even in this much simpler case, there are still many open
questions.  I want to illustrate this fact for the example in which the LSND
signal is ignored.

In view of the strong CH00Z bound on $\nu_e\to\nu_X$ oscillations, if one
ignores the LSND signal one can contemplate rather separate squared mass
difference for the atmospheric and solar case.  Taking $\Delta m^2_{\rm atmos} =
\Delta m^2_{23} \sim 10^{-3}~{\rm eV}^2$ and $\Delta m^2_{\rm solar} =
\Delta m^2_{12}$, then the CH00Z result~\cite{CH00Z} implies
$\theta_{13}\simeq 0$.  Neglecting any CP violating effect---a reasonable
first approximation---then the neutrino mixing matrix which describes the
observed phenomenon is given approximately by
\begin{eqnarray}
U &\simeq &\left[
\begin{array}{ccc}
1 & 0 & 0 \\
0 & 1/\sqrt{2} & -1/\sqrt{2} \\
0 & 1/\sqrt{2} & 1/\sqrt{2}
\end{array} \right]
\left[
{\bf{1}} \right]
\left[
\begin{array}{ccc}
c_{12} & -s_{12} & 0 \\
s_{12} & c_{12} & 0 \\
0 & 0 & 0
\end{array} \right] \nonumber \\
&=&
\left[
\begin{array}{ccc}
c_{12} & -s_{12} & 0 \\
s_{12}/\sqrt{2} & c_{12}/\sqrt{2} & -1/\sqrt{2} \\
s_{12}/\sqrt{2} & c_{12}/\sqrt{2} & 1/\sqrt{2}
\end{array} \right]~.
\end{eqnarray}
That is, there is maximal mixing for atmospheric neutrinos
($\theta_{23}\simeq 45^o$), arbitrary solar mixing ($\theta_{12}$), and
$\theta_{13}\simeq 0^o$.

Even given the above mixing matrix, many questions remain.  For instance,
is maximal mixing allowed, $\theta_{12}\simeq 45^o$?  Is a totally
degenerate neutrino mass spectrum $(m_1=m_2=m_3 \simeq 0.5~{\rm eV})$
allowed?  What is the origin of the neutrino mass matrix $M_\nu$ which has
$\theta_{23}\simeq 45^o$ but $\theta_{12},~\theta_{13}\simeq 0$?  More
generally, since the neutrino mass matrix $M_\nu$ most probably results
from a see-saw mechanism with
\begin{equation}
M_\nu = m_{\rm D}^T m_{\rm S}^{-1}m_{\rm D}
\end{equation}
what are the natural Dirac matrices $m_{\rm D}$ and Majorana matrices
$m_{\rm S}$ which produce the mixing matrix $M_{\nu}$?  
These questions and others are
addressed in different ways in the burgeoning literature on neutrino
masses,~\cite{literature} but progress most probably must await further
experimental input.

\section{Hopes}
\subsection{Windows of Opportunity}

As just remarked, it is clear that progress in understanding what is going
on in the neutrino sector can only come from further data.  Fortunately,
new data will be forthcoming in all relevant $\Delta m^2$ ranges.  For solar
neutrinos, in the next five years, three experiments (SNO, Borexino and
SuperKamiokande) should be able to definitely answer the question of whether
matter induced (MSW) oscillations are involved. SNO, in particular, 
measures both NC $(\nu_Xd\to np\nu_X)$ and CC $(\nu_ed\to ppe^-)$ processes
and their results should 
be free of solar model ambiguities.  Also Borexino, by being mainly 
sensitive to neutrinos produced by the Be reactions, is crucially sensitive to whether one has , or has not, matter induced
oscillations in the sun.

Similar progress is expected in the atmospheric neutrino front.  Here
long-baseline experiments will probe the $\Delta m^2-\sin^22\theta$ region
identified by SuperKamiokande with accelerator neutrino beams. Typically, ~\cite{Conrad} the planned experiments at Fermilab and CERN will
cover the region $\Delta m^2\stackrel{>}{_{\scriptstyle \sim}}
10^{-3}$ down to $\sin^22\theta$ of order $10^{-1}$.  Thus they should
be able to handily confirm the SuperKamiokande signal and begin to explore
its nature in some more detail.  There is also a planned experiment at
Fermilab [MiniBoone] which will cover the $\Delta m^2-\sin^22\theta$ region
for $\nu_e\to\nu_\mu$ oscillations identified by LSND with nearly an order of
magnitude more sensitivity.\cite{Conrad}

These neutrino ``windows of opportunity" are going to be parallelled in the
quark sector.  We shall soon have 
results on $\epsilon^\prime/\epsilon$ from KTeV and
NA48 and hopefully finally learn that there is direct CP violation, as
predicted by the CKM model.  However, the crucial tests for this model
awaits the turn-on of the SLAC and KEK B factories in 1999 and the measurements
of CP-violating processes in the B system.  It was quite clear in Vancouver
that the community is chomping at the bit to start doing this physics!  For instance, both OPAL~\cite{Barberio} and
CDF~\cite{Tseng} presented first attempts at extracting $\sin2\beta$.  
Although the numbers obtained are not statistically significant, it was nice
to see that the systematic error is beginning to be under some control
[$\pm 0.5$ for OPAL and $\pm 0.3$ for CDF].  Indeed, CDF was able to exclude
at the 95\% confidence level values of $\sin2\beta < -0.20$.\cite{Tseng}

The expectation of the CKM model is that $\sin2\beta$ is large and positive.
The CKM matrix analyses of Parodi~\cite{Parodi} quoted earlier, gives for
this parameter $\sin 2\beta = 0.73 \pm 0.08$. However,
the other angles in the unitarity triangle are less well determined
(e.g. in Parodi's analysis $\sin2\alpha = -0.15\pm 0.30$ and
$\gamma = (62 \pm 10)^o$).  Thus to check the CKM model in detail in the B
factories (and elsewhere) will be a challenging task.  The magnitude of this
challenge was made apparent in Alexander's talk~\cite{Alexander} who reported
on the branching ratios of $B_d$ into $K^\pm\pi^\mp$ and $\pi^\pm \pi^\mp$.
Although the $K\pi$ branching ratio is now well established~\cite{Alexander}
[${\rm BR}(B_d\to K^\pm\pi^\mp) = (1.4 \pm 0.3 \pm 0.1)\times 10^{-5}$],  the $\pi\pi$ branching rate is not yet seen at a
significant level.  Since the decay $B_d\to\pi\pi$ is one of the main
ways to get at $\sin 2\alpha$, pinning down soon this decay rate is of
considerable importance.

Experiments with neutrinos and $B$-decays are likely to rivet the attention
of our field in the near future, making the next few years the
{\bf golden flavor years}.  Nevertheless, one should not forget that there
are also important opportunities ahead for probing the physics of the
weak scale, {\bf before} the turn-on of the LHC.  As McNamara~\cite{McNamara}
and Treille~\cite{Treille} discussed, LEP200 has still about 10 GeV of energy
range to explore, with good luminosity.  This should make it possible to
find the Higgs boson if its mass is less than 105 GeV---a good 15 GeV above
the present bound.  This reach is shown in Figure 5.

\begin{figure}
\center
\epsfig{file=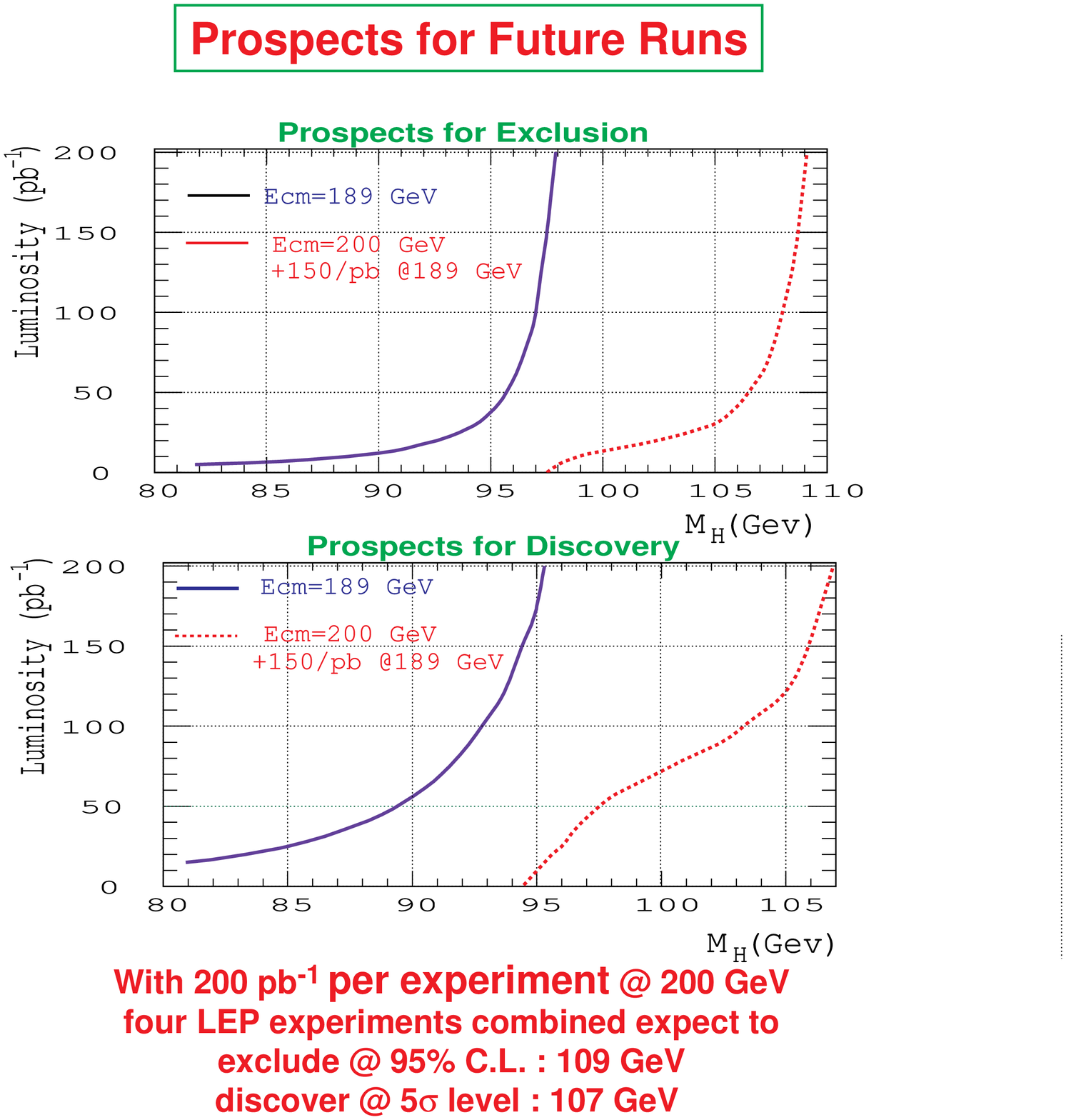,height=4in}
\caption{Prospects for discovering the Higgs at LEP200.}
\end{figure}

There  are also ample opportunities at the Fermilab Tevatron, which should soon be running with the
new Main Injector.  The Tevatron/MI will have higher luminosity, which should
allow it to collect 2 $fb^{-1}$ in Run II.  Many searches now at the
Collider have, typically, 4 events with a background of 3 events.  A factor of
20 increase in luminosity in these circumstances can work wonders!  Furthermore,
the luminosity for the Tevatron/MI should increase further beyond Run II.
If the Fermilab collider can increase its integrated luminosity to around
$\int {\cal{L}} dt \sim (20-25)~fb^{-1}$, then it might be able to have a
shot at Higgs masses as high as $M_{\rm H} = 120,~{\rm GeV}$~\cite{Valls,Treille} before the turn-on of the LHC.

\subsection{The Long Road Ahead.}

Around 2005 we will begin the detailed exploration of the physics of
electroweak symmetry breaking at the LHC.  This will be an important milestone
for high energy physics and one which particle physicists have had their
eyes on for a long time.  Even so, it is the nature of our field to 
already plan both machines which are {\bf complementary} to the LHC---like the
$e^+e^-$ linear collider---and machines which {\bf go beyond} the LHC---like
the muon collider and the very large hadron collider.  Indeed at ICHEP 98
there was a whole parallel session devoted to future machines. This was also the focus of  Kurt Hubner's~\cite{Hubner} plenary presentation.

I do not want to comment in any detail on these future prospects here.
However, I would like to make a general observation, which I hope will be
helpful.  All these future prospects face enormous technical, political and
economic challenges, which are really intrinsic for projects of this
magnitude.  To have any chance of success, in my view, it is important  that
we act as a coherent community internationally.  Furthermore, since
ultimately the sources of funding for these projects are a tax on our
fellow citizens, we have an obligation to do our utmost to explain and 
popularize what we do for the benefit of the general public.
\vskip.5cm

\section*{Acknowledgments}
\vskip.3cm

I am very grateful to all the scientific secretarial staff at ICHEP 98 in
Vancouver.  They were enormously helpful to me during the preparation of this
talk and transformed a daunting task into (almost!) a delight.  This work
is supported in part by the Department of Energy under Contract No. 
DE-FG03-91ER40662. Task C.

\end{document}